\documentclass[a4paper,11pt]{article}
\pdfoutput=1

\usepackage{layaureo}   %better a4 space usage, but less violent than a4wide	
\usepackage{hyperref}   %links and urls
\usepackage{natbib}     %author-year citation support
\usepackage{url}

\usepackage{booktabs}      %better tables
\usepackage{csvsimple-l3}  %read data from csv files
   
\usepackage{graphicx}   %to instert images

\usepackage{bm}
\usepackage{amsmath}    %standard utils
\usepackage{mathtools}  %extension of the above
\usepackage{amsfonts}	%more fonts
\usepackage{amssymb}	%more symbols
\usepackage{amsthm}     %standard environments
\usepackage{mathrsfs}   %better curly fonts
\usepackage{cancel}     %semplifications

\usepackage[capitalize]{cleveref}

\theoremstyle{plain}
\newtheorem{thm}{Theorem}[section]

\theoremstyle{definition}

\newtheorem{hp}[thm]{Hypothesis}

\newtheorem{exmp}[thm]{Example}

\theoremstyle{remark}
\newtheorem{rem}[thm]{Remark}

\numberwithin{equation}{section}

  {\left\lbrace\begin{array}{@{}l@{}}}%
  {\end{array}\right.}

\newcommand{\R}{\mathbb{R}}

\newcommand{\PR}{\mathbb{P}}
\newcommand{\E}{\mathbb{E}}

\newcommand{\eps}{\varepsilon}

\newcommand{\der}[2]{\frac{\partial{#1}}{\partial{#2}}} %derivata parziale (2 argomenti)
\newcommand{\derXX}[2]{\frac{\partial^2{#1}}{\partial{#2}^2}} %derivata seconda (2 argomenti)
\newcommand{\derXY}[3]{\frac{\partial^2{#1}}{\partial{#2}\partial{#3}}} %derivata seconda mista (3 argomenti)
\newcommand{\ider}[2]{\ifrac{\partial{#1}}{\partial{#2}}} %versione inline
\newcommand{\iderXX}[2]{\ifrac{\partial^2{#1}}{\partial{#2}^2}} %versione inline
\newcommand{\iderXY}[3]{\ifrac{\partial^2{#1}}{\partial{#2}\partial{#3}}} %versione inline
\newcommand{\tr}[1]{{#1}^{\intercal}} %transpose
\newcommand{\diff}{\mathrm{d}} %differential
\newcommand{\vc}[1]{{\bm #1}} %abstraction for vector (font choice: bold)
\newcommand{\mt}[1]{\mathrm{\mathbf{#1}}} %abstraction for matrix (font choice: straight bold)
\newcommand{\ifrac}[2]{#1/#2} %fraction in inline contexts

\newcommand{\quoted}[1]{``#1''} %quotation marks

\newcommand{\NPV}{\ensuremath{\mathrm{NPV}}}

\newcommand{\lgd}{\ensuremath{\mathrm{lgd}}}
\newcommand{\aalpha}{\vc{\alpha}}
\newcommand{\zzero}{\vc{0}}

\title{Fast and Stable Credit Gamma of CVA}
\author{Roberto Daluiso\thanks{Interest Rates and Credit Models, IMI Corporate and Investment Banking, Intesa Sanpaolo. E-mail: \texttt{roberto.daluiso@intesasanpaolo.com}.}}
\date{First version: November 20, 2023}
\providecommand{\keywords}[1]{\small\textbf{Keywords:} #1}

\begin{document}

\maketitle

\begin{abstract}
Credit Valuation Adjustment is a balance sheet item which is nowadays subject to active risk management by specialized traders.
However, one of the most important risk factors, which is the vector of default intensities of the counterparty, affects in a non-differentiable way the most general Monte Carlo estimator of the adjustment, through simulation of default times.
Thus the computation of first and second order (pure and mixed) sensitivities involving these inputs cannot rely on direct path-wise differentiation, while any approach involving finite differences shows very high statistical noise.
We present \textit{ad hoc} analytical estimators which overcome these issues while offering very low runtime overheads over the baseline computation of the price adjustment.
We also discuss the conversion of the so-obtained sensitivities to model parameters (e.g.~default intensities) into sensitivities to market quotes (e.g.~Credit Default Swap spreads).
\end{abstract}

\keywords{Valuation Adjustments, Monte Carlo, derivatives pricing, algorithmic differentiation, adjoints, Greeks, sensitivities, Hessian, copula.}

\section{Introduction}\label{sec:intro}

When a counterparty defaults on an in-the-money portfolio, the surviving party experiences a loss. Modern risk management and accounting standards dictate to adjust the portfolio value by the risk-neutral expected value of such loss, called Credit Valuation Adjustment \citep[CVA; see e.g.][]{BrigoEtAl2013ctpyCollateralFunding}.\footnote{The full paper would also apply to the completely analogous Debt Valuation Adjustment (DVA), since it is defined as the CVA seen by the counterparty on the same position.}
Daily changes in CVA contribute to the the Profit and Loss, and are therefore managed by dedicated trading desks.

The universally accepted model for the loss is a fraction of the positive value of the portfolio at default time.
The positive part in such model introduces a non-linearity that makes CVA a portfolio-wise metric, which in general can only be computed by a costly Monte Carlo simulation with tens or hundreds of inputs.
Therefore, even the simplest representation of risks by first order sensitivities is a major computational challenge, while a second order representation is usually considered out of reach.

As for the efficient computation of first order derivatives, a partial solution has been found in the adjoint algorithmic differentiation (AAD) technique, which gives the full gradient of any smooth deterministic function with at most a 4x overhead over the original computational time, regardless of the number of inputs \citep{GriewankWalther2008evaluatingDerivatives}.
This can be leveraged to estimate expected values of payoffs computed by Monte Carlo, by path-wise computation of the derivatives \citep{Glasserman2004mcMethods}, but only when the payoff is expressed as a differentiable function of the inputs.
In particular, if CVA is implemented directly by simulation of defaults, then path-wise AAD often covers the sensitivities to model parameters entering only the portfolio value, but it does not apply to sensitivities to credit model parameters, because of the discontinuous dependence on default time.
To our knowledge, this issue has been explicitly addressed only in \citet{CapriottiEtAl2011realTimeCtpyCreditRiskInMC} by a discrete-time approximation of the CVA payoff.
On the other hand, out of the CVA setting, several ways to apply AAD to discontinuous payoffs have been proposed in the literature \citep{Giles2007mcSens, ChanJoshi2015optimalLimitMethods, DaluisoFacchinetti2018adDiscontinuous}, but mostly in diffusive settings which do not cover default simulation.

Second order derivatives have been studied only for simpler problems than CVA \citep{Capriotti2015secondOrder, PagesEtAl2016adHigherOrder, JoshiZhu2016optimalPartialProxyGammas, Daluiso2020secondOrder}.
We refer interested readers to the latter, which includes a detailed comparison with all the others in a diffusive setting.
For our purposes, the key takeaways are that on the one hand, in most cases the best trade-off between per-path runtime and statistical noise is a method whose cost is linear in the number of variables; but on the other hand, constant-overhead estimators exist, and for some forms of the payoff they display spectacular efficiency.

Finally, the above algorithms produce sensitivities to model parameters; however, to hedge the measured risk, the trader needs a rule to translate this information into sensitivities to market instruments entering the calibration of models.
It is known that first order derivatives can be converted into hedge ratios by very fast implicit-function based algorithms \citep{Henrard2011ADimplicitFunction, Henrard2013ADleastSquare, Daluiso2016modelToMarket}; while as far as we know, the conversion of model Hessians to second order Greeks has never been discussed.

Given the state of the art described above, the contribution of this paper is threefold.

Firstly, we show how the main approaches to first-order discontinuous AAD can adapted to payoffs depending on default time, like that appearing in the CVA definition.
We motivate our choice of one of them and test it numerically, getting excellent results in terms of both speed and accuracy. 

Secondly, we tackle second order differentiation concentrating on cases in which the set of differentiation variables can be split in two parts: the first one $\vc{\psi}$ includes market inputs not affecting the default time, while the complementary $\vc{\theta}$ affects only the default model. This applies to the CVA of portfolios which do not include credit derivatives. For such common setup, we introduce an estimator of $\iderXY{}{\vc{\theta}}{\vc{\psi}}$ not slower than a constant multiple of the pricing-only run time; and an analogous estimator for $\iderXX{}{\vc{\theta}}$ for which such overhead is also constant up to a practically negligible add-on.
We present numerical tests of both, in which we observe empirical gains in uncertainties and performance which are even more impressive than for the first order.

Thirdly, we derive formulas to convert second order sensitivities to model parameters into second order sensitivities to market quotes.
In such respect, we provide both a fully analytical approach generalizing the first order literature mentioned above, and a practical alternative, partly based on finite differences.

The rest of the paper is organized as follows.
\Cref{sec:setting} formalizes the problem of interest and introduces the main notation.
\Cref{sec:delta,sec:gamma} derive our estimators respectively for first and second order model sensitivities.
\Cref{sec:conversion} presents the conversion into sensitivities to market quotes.
\Cref{sec:numerics} tests numerically the proposals of \cref{sec:delta,sec:gamma}.
A summary of the main findings is eventually provided in \cref{sec:conclusion}.

\section{Setting}\label{sec:setting}

We consider an expected value depending on a vector of parameters $\vc{\alpha}$
\[
	p(\aalpha) = \E[f(\vc{\tau},\vc{X},\vc{\alpha})],
\]
where:
\begin{itemize}
\item $\vc{X}$ is a generic stochastic driver (e.g.~a multi-dimensional standard Brownian motion).
\item $\vc{\tau}$ is a vector of default times $\tau^{(i)}$ for $i=1,\dots,\bar{\imath}$ driven by default intensities $\lambda^{(i)}(\cdot,\vc{X},\aalpha) > 0$, possibly dependent on the stochastic driver $\vc{X}$:
\begin{equation*}
\PR(\tau^{(i)} > t \mid \vc{X}) = e^{-\Lambda^{(i)}(t,\vc{X},\aalpha)},\quad  \Lambda^{(i)}(t,\vc{X},\aalpha):=\int_0^t \lambda^{(i)}(s,\vc{X},\aalpha)\, \diff{s}.
\end{equation*}
When $\vc{\tau}$ is unidimensional, we drop the superscripts of $\tau$, $\lambda$ and $\Lambda$.
\item $f$ is a deterministic map representing the payoff, typically inclusive of the simulation of market drivers as a smooth function of the parameters $\vc{\alpha}$ and random draw $\vc{X}$ (e.g.~by Euler discretization of a Stochastic Differential Equation).
\end{itemize}

Our first aim is to compute first and second order derivatives of $p$ with respect to $\vc{\alpha}$ by Monte Carlo. We suppose for simplicity that the direct dependence of $f$ on $\vc{\alpha}$ is differentiable. On the other hand, we do not assume differentiability with respect to $\vc{\tau}$. Indeed, in the main application we have in mind, which is CVA, we have
\begin{equation}\label{eq:ucva}
	p = -\lgd\cdot\E\left[D(0,\tau)(\NPV_\tau)^{+}I_{\{\tau \leq T\}}\right]
\end{equation}
for unilateral computations neglecting own default, or
\begin{equation}\label{eq:bcva}
	p = -\lgd\cdot\E\left[D(0,\tau^{(1)})(\NPV_{\tau^{(1)}})^{+}I_{\{\tau^{(1)} \leq \min(T,\tau^{(2)})\}}\right]
\end{equation}
for bilateral computations: in such expressions, the function $\NPV_s$ has a jump at all times $s$ in which a payment occurs, $\lgd$ and $T$ are constants, and $D(0,\cdot)$ is a deterministic discounting function.

Conversion formulas will be also provided for the case in which $\aalpha$ is in fact only a convenient intermediate deterministic function of the real inputs of interest: see \cref{sec:conversion} for the specific framework.

\section{First order model sensitivities}\label{sec:delta}

In this section we focus on the estimation of $\ider{p}{\vc{\alpha}}$.

\subsection{By distributional differentiation}\label{sec:distributional}

In this subsection we suppose for simplicity that $\bar{\imath}=1$, i.e.~only one default time enters the payoff; estimators allowing for multiple defaults will be covered in \cref{sec:multiple}.

The most natural idea is probably to consider $f(\vc{X},\vc{\alpha},\tau)$ as a function 
\[
	\phi(\eps,\vc{X},\vc{\alpha}) := f(\Lambda(\cdot,\vc{X},\aalpha)^{-1}(\eps),\vc{X},\vc{\alpha})
\]
of $\vc{\alpha}$ inclusive of the well-known simulation scheme of $\tau$ from an exponential draw $\eps$ independent of $\vc{X}$. Under this point of view, one can just adapt one of the methods for AAD with discontinuous payoffs.

In a diffusive context, \citet{DaluisoFacchinetti2018adDiscontinuous} show that one with excellent efficiency can be obtained by differentiating indicator functions in distributional sense:
\begin{equation*}
	\der{}{\aalpha}I_{f(\aalpha,\vc{z})>0}=\der{}{f}I_{f>0}\cdot\der{f}{\aalpha}(\aalpha,\vc{z})=\left[\delta(f)\der{f}{\aalpha}\right](\aalpha,\vc{z}),
\end{equation*}
where $\delta$ is the Dirac delta distribution. Therefore, we adapt their heuristic argument to the dependence on the non-Gaussian $\eps$. To this purpose, we suppose that the dependence of $f$ on $\tau$ is differentiable everywhere except at a finite set of times $T_i$ where it has jumps $\Delta_i(\vc{X},\aalpha)$, as in virtually every financial application. Then we get
\begin{align*}
	\der{p}{\aalpha} &= 
	\E\left\{ \der{\phi}{\aalpha} + \sum_i \Delta_i \delta(\tau - T_i) \der{\tau}{\aalpha} \right\}
	= \E\left\{ \der{\phi}{\aalpha} - \sum_i \Delta_i\left[\lambda e^{-\Lambda}\right] \left[\frac{1}{\lambda}\der{\Lambda}{\aalpha} \right]\right\}\\
	& = \E\left\{ \der{\phi}{\aalpha} - \sum_i \Delta_ie^{-\Lambda}\der{\Lambda}{\aalpha}\right\},
\end{align*}
where the first quantity in square brackets is the known density of $\tau$, and the second one is the derivative of the inverse function $\Lambda(\cdot,\vc{X},\aalpha)^{-1}$.

A more formal derivation of the same result is as follows, where we define conventionally $T_0=0$ and $T_{\bar{\imath}+1}=\infty$ and use the notation $\phi(u+,\vc{X},\aalpha)$ and $\phi(u-,\vc{X},\aalpha)$ to denote left and right limits of $\phi$:
\begin{multline*}
\der{}{\aalpha} \E\left[ \phi(\eps,\vc{X},\vc{\alpha}) \right] = 
 \der{}{\aalpha} \E\left[ \sum_{i = 0}^{\bar{\imath}} \int_{\Lambda(T_i,\vc{X},\aalpha)}^{\Lambda(T_{i+1},\vc{X},\aalpha)} \phi(u,\vc{X},\vc{\alpha})e^{-u}\,\diff{u} \right] \\
= \E\left[\sum_{i=0}^{\bar{\imath}} \left(\phi(\Lambda(T_{i+1},\vc{X},\aalpha)-,\vc{X},\vc{\alpha})e^{-\Lambda(T_{i+1},\vc{X},\aalpha)}\der{\Lambda(T_{i+1},\vc{X},\aalpha)}{\aalpha} + \right. \right. \\
\left. \left . - \phi(\Lambda(T_{i},\vc{X},\aalpha)+,\vc{X},\vc{\alpha})e^{-\Lambda(T_{i},\vc{X},\aalpha)}\der{\Lambda(T_{i},\vc{X},\aalpha)}{\aalpha}\right) + \int_0^{\infty} \der{\phi}{\aalpha}(u,\vc{X},\vc{\alpha})e^{-u}\,\diff{u} \right]\\
= \E\left[\der{\phi}{\aalpha}(\eps,\vc{X},\vc{\alpha}) - \sum_{i=1}^{\bar{\imath}} \left( f(T_i+,\vc{X},\vc{\aalpha}) - f(T_i-,\vc{X},\vc{\aalpha}) \right)e^{-\Lambda(T_{i},\vc{X},\aalpha)} \der{\Lambda(T_{i},\vc{X},\aalpha)}{\aalpha} \right].
\end{multline*} 
The validity of the above needs only that all the derivatives which appear in the formulas exist, and that we can exchange integrals with differentiation. 

\begin{rem}[Algorithmic complexity]\label{rem:distributional_complexity}
The correction term $\Delta_i\exp{(-\Lambda)}\ider{\Lambda}{\aalpha}$ appears $\bar{\imath}$ times, one for each discontinuity $T_i$: in the CVA case, at each cash-flow date. Therefore, the raw computational cost per Monte Carlo path is unacceptably high for large portfolios. A more scalable algorithm is presented in the next subsection.
\end{rem}

\subsection{By conditional differentiation}\label{sec:conditional}

This subsection keeps the assumption that $\bar{\imath}=1$, again deferring multi-default generalizations to \cref{sec:multiple}.

Here instead of considering $\tau$ as a deterministic function of $\aalpha$ and of random draws not depending on $\aalpha$, we consider $\tau$ as a random variable and consider the dependence of its $\vc{X}$-conditional \emph{law} from $\aalpha$.

The result is a correction to the path-wise estimator reminiscent of the likelihood ratio method \citep{Glasserman2004mcMethods}, but applied to a conditional distribution. We can therefore hope that it has lower Monte Carlo uncertainties than plain likelihood ratios, analogously to what \citet{Giles2007mcSens} finds in purely diffusive settings by conditioning on the full discretized path with only one point excluded.

The derivation is again based on differentiation under expectation and integral:
\begin{align*}
\der{p}{\aalpha} &= \der{}{\aalpha} \E\biggl\{ \E\left[ f(\tau,\vc{X},\aalpha) \mid \vc{X} \right] \biggr\} =
\der{}{\aalpha} \E\biggl\{ \int_0^{\infty} \left[fe^{-\Lambda}\lambda\right](t,\vc{X},\aalpha)\,\diff{t} \biggr\}\\
&=\E\biggl\{ \int_0^{\infty} \left[\biggl( \der{f}{\aalpha} + f\frac{(\partial/\partial\aalpha)(e^{-\Lambda}\lambda)}{e^{-\Lambda}\lambda} \biggr)
e^{-\Lambda}\lambda\right](t,\vc{X},\aalpha)\,\diff{t} \biggr\},
\end{align*}
where we factored out the conditional density $e^{-\Lambda}\lambda$ so that the result can be expressed again as an expected value:
\begin{equation}\label{eq:conditional}
\der{p}{\aalpha}
= \E\biggl\{ \left[\der{f}{\aalpha} + f\der{}{\aalpha}\log\left(e^{-\Lambda}\lambda\right)\right](\tau,\vc{X},\aalpha) \biggr\}
= \E\biggl\{ \left[\der{f}{\aalpha} + f\der{w}{\aalpha}\right](\tau,\vc{X},\aalpha) \biggr\}
\end{equation}
with the definition 
\[
	w(\tau,\vc{X},\aalpha) := -\Lambda(\tau,\vc{X},\aalpha) +\log\lambda(\tau,\vc{X},\aalpha).
\]

\begin{rem}[Algorithmic complexity]\label{rem:conditional_complexity}
Equation \eqref{eq:conditional} can be implemented by computing a single gradient on each Monte Carlo path: by AAD, this has a very low overhead over the computation of $f$ as explained in the introduction.
\end{rem}

\subsection{With multiple defaults}\label{sec:multiple}

In this subsection we drop the hypothesis that $f$ depends only on a single default time $\tau$.

We first present the case in which the default triggers
\[
	\eps^{(i)} := \Lambda^{(i)}\left(\tau^{(i)},\vc{X}, \aalpha\right)
\]
are independent. Note that this does not mean that single name default times are independent, as default intensities may be correlated, and/or some of them may represent joint defaults \textit{à la} Marshall-Olkin \citep[see e.g.][]{Morini2011modelRisk}.

With this assumption,
\[
	\PR\left(\tau^{(i)}>t_i \ \forall i=1,\dots,\bar{\imath} \mid \vc{X}\right) = \prod_{i=1}^{\bar{\imath}} \exp\left(-\Lambda^{(i)}\left(t_i, \vc{X}, \aalpha\right)\right),
\]
so by the same ideas as in \cref{sec:conditional}
\begin{align*}
	\der{p}{\aalpha}
	&= \der{}{\aalpha} \E\biggl\{ \E\left[ f\left(\vc{\tau}, \vc{X}, \aalpha\right) \mid \vc{X} \right] \biggr\}\\
	&= \der{}{\aalpha} \E\biggl\{ \int_{\R_+^{\bar{\imath}}} f\left(\vc{t},\vc{X},\aalpha\right)\prod_{i=1}^{\bar{\imath}} \left[\exp\left(-\Lambda^{(i)}\right)\lambda^{(i)}\right]\left(t_i, \vc{X}, \aalpha\right)\, \diff{\vc{x}}\biggr\}\\
	&= \E\biggl\{ \der{f}{\aalpha}\left(\vc{\tau}, \vc{X}, \aalpha\right) + f\left(\vc{\tau}, \vc{X}, \aalpha\right)\der{}{\aalpha}\log\prod_{i=1}^{\bar{\imath}} \left[\exp\left(-\Lambda^{(i)}\right)\lambda^{(i)}\right](\tau^{(i)}, \vc{X}, \aalpha) \biggr\}
\end{align*} 
which takes the same form of the unidimensional estimator
\begin{equation}\label{eq:multiple}
\der{p}{\aalpha}
= \E\biggl\{ \left[\der{f}{\aalpha} + f\der{w}{\aalpha}\right](\vc{\tau},\vc{X},\aalpha) \biggr\}
\end{equation}
with the generalized definition 
\begin{equation}\label{eq:log_density_indep}
	w(\vc{\tau},\vc{X},\aalpha) := \sum_{i=1}^{\bar{\imath}} \left[-\Lambda^{(i)} +\log\lambda^{(i)}\right](\tau^{(i)}, \vc{X}, \aalpha).
\end{equation}

In general, the $\vc{X}$-conditional dependence among the exponentially distributed default triggers can be expressed by a copula $C_{\vc{X},\aalpha}$. Conditional on $\vc{X}$, $\tau^{(i)}$ is a monotone transformation of $\eps^{(i)}$, so $C_{\vc{X},\aalpha}$ is also the $\vc{X}$-conditional copula of default times. 

Let us suppose that the copula $C_{\vc{X},\aalpha}$ has a density $\rho_{\vc{X},\aalpha}$ with respect to the Lebesgue measure. In such case, it is well known that $\vc{\tau}$ also has a density, given by the product of $\rho_{\vc{X},\aalpha}$ with the density of the marginals: specifically, defining the $\vc{X}$-conditional grades
\[
	\vc{u}(\vc{t},\vc{X},\vc{\alpha}) := \left(1-e^{-\Lambda^{(1)}(t^{(1)}, \vc{X}, \aalpha)},\dots,1-e^{-\Lambda^{(\bar{\imath})}(t^{(\bar{\imath})}, \vc{X}, \aalpha)}\right)
\]
one has
\[
	\PR\left( \vc{\tau}\in\diff{}\vc{t} \mid \vc{X} \right) =  \rho_{\vc{X},\aalpha}\left(\vc{u}(\vc{t},\vc{X},\vc{\alpha})\right)\prod_{i=1}^{\bar{\imath}} \left[\exp\left(-\Lambda^{(i)}\right)\lambda^{(i)}\right]\left(t_i, \vc{X}, \aalpha\right)\,\diff{}\vc{t}.
\]

This means that the derivation leading to \eqref{eq:multiple} is still valid with an additive correction to the log-density \eqref{eq:log_density_indep}:
\begin{equation}\label{eq:log_density_copula}
	w(\vc{\tau},\vc{X},\aalpha) := \log \rho_{\vc{X},\aalpha}\left(\vc{u}(\vc{\tau},\vc{X},\vc{\alpha})\right) + \sum_{i=1}^{\bar{\imath}} \left[-\Lambda^{(i)} +\log\lambda^{(i)}\right](\tau^{(i)}, \vc{X}, \aalpha).
\end{equation}
One can also note that
\[
	\der{}{\aalpha}\bigg[\log \rho_{\vc{X},\aalpha}\left(\vc{u}(\vc{\tau},\vc{X},\vc{\alpha})\right)\bigg] =
	\der{\log\rho_{\vc{X},\aalpha}}{\aalpha} + \sum_{i=1}^{\bar{\imath}} \der{\log\rho_{\vc{X},\aalpha}}{u_i}(1-u_i)\der{\Lambda^{(i)}}{\aalpha}(\vc{\tau},\vc{X},\vc{\alpha})
\]
and therefore a different definition of $w(\vc{\tau},\vc{X},\aalpha)$ yielding the same \eqref{eq:multiple} is
\begin{equation}\label{eq:log_density_copula_alternative}
	\log\rho_{\vc{X},\aalpha}(\vc{u}) + \sum_{i=1}^{\bar{\imath}} \left[ \left(d_i-1\right)\Lambda^{(i)} +\log\lambda^{(i)}\right](\tau^{(i)}, \vc{X}, \aalpha)
\end{equation}
where the definition of the constants $d_i$
\begin{equation}\label{eq:constants}
	d_i := \der{\log \rho_{\vc{X},\aalpha}}{u_i}(\vc{u})(1-u_i)
\end{equation}
should not be differentiated, nor the argument $\vc{u}$ of $\rho_{\vc{X},\aalpha}$.

\begin{rem}[Algorithmic complexity]
As in \cref{rem:conditional_complexity}, the computational cost of estimator \eqref{eq:multiple} is a small multiple over that of the Monte Carlo estimation of $p$ thanks to AAD.
\end{rem}

\subsection{With censored default times}\label{sec:censored}

In practical implementations, default happening after a finite time $T$ may be neglected as irrelevant, and therefore default intensities may be simulated only up to time $T$; this is the case for CVA, where $T$ is the maturity of the portfolio. This means that if the set of defaulted names is $\mathcal{I}\subseteq\{1,\dots,\bar{\imath}\}$ and $\mathcal{I}^c$ is its complementary, we only know $\vc{\tau}^{(\mathcal{I})} := (\tau^{(i)})_{i\in\mathcal{I}}$, while for $j\notin \mathcal{I}$ we only know that $\tau^{j}>T$. So \eqref{eq:log_density_indep}-\eqref{eq:log_density_copula} cannot be computed as they are; this subsection describes techniques to solve this issue.

\subsubsection{Keeping samples from the full copula}

A first way to derive an estimator in this setting is to notice that by hypothesis, the following modified intensity model gives the same payoff as the original model:
\begin{equation}\label{eq:censored_intensity}
	\tilde{\lambda}^{(i)}(t,\vc{X},\vc{\alpha}) = \lambda^{(i)}(t,\vc{X},\vc{\alpha})I_{t\leq T} + \tilde{\lambda}^{(i)}_{\infty}I_{t>T},
\end{equation}
where $\tilde{\lambda}^{(i)}_{\infty}>0$ is an arbitrary constant which will drop out of the final result. Under this model we have
\begin{align*}
	\der{\tilde{\Lambda}^{(i)}}{\aalpha}(\tilde{\tau}^{(i)}, \vc{X}, \aalpha) &= \der{\Lambda^{(i)}}{\aalpha}(\tau^{(i)}\wedge T, \vc{X}, \aalpha),\\
	\der{\tilde{\lambda}^{(i)}}{\aalpha}(\tilde{\tau}^{(i)}, \vc{X}, \aalpha) &= I_{\{\tau^{(i)}\leq T\}}\der{\lambda^{(i)}}{\aalpha}(\tau^{(i)}, \vc{X}, \aalpha),
\end{align*}
so we can modify \eqref{eq:log_density_copula_alternative} as follows without affecting the gradient $\ider{w}{\aalpha}$ in \eqref{eq:multiple}:
\begin{equation}\label{eq:log_density_censored_intensity}
	\log\rho_{\vc{X},\aalpha}(\vc{u}) + \sum_{i=1}^{\bar{\imath}} \left[ \left(d_i-1\right)\Lambda^{(i)}(\tau^{(i)}\wedge T, \vc{X}, \aalpha) + I_{\{\tau^{(i)}\leq T\}}\log\lambda^{(i)}(\tau^{(i)}, \vc{X}, \aalpha)\right].
\end{equation}
This formula can be computed without knowing the exact default time $\tau^{(j)}$ for $j\notin\mathcal{I}$, although in general it does need $u_j$; this should not be problematic, as one should have sampled the full vector $\vc{u}$ anyway, to know both the exact $\tau^{(i)}$ for $i\in\mathcal{I}$ and that $u_j > 1-\exp(-\Lambda^{(j)}(T))$ for $j\notin\mathcal{I}$. However, if default triggers are independent, all terms involving $\vc{u}$ vanish and we get
\begin{equation}\label{eq:log_density_censored_indep}
	\sum_{i=1}^{\bar{\imath}} \left\{I_{\{\tau^{(i)}\leq T\}} \left[-\Lambda^{(i)} +\log\lambda^{(i)}\right](\tau^{(i)}, \vc{X}, \aalpha) - I_{\{\tau^{(i)}> T\}}\Lambda^{(i)}(T, \vc{X}, \aalpha) \right\}.
\end{equation}

\subsubsection{Integrating out or subtracting survived names}

Without independence, an estimator in which default triggers of surviving names do not appear can also be derived, but its implementation needs more tractability of the copula. To obtain it, we define the partition of $\R_{+}^{\bar{\imath}}$
\[
	E_{\mathcal{I}} := \left\{\vc{t} \colon \left(t_i \leq T \; \forall i \in \mathcal{I}\right) \wedge \left(t_j > T \; \forall j \in \mathcal{I}^c\right)\right\}, \qquad \mathcal{I}\subseteq\{1,\dots,\bar{\imath}\},
\]
and the conditional probabilities
\[
	\pi_{\mathcal{I}}\left(\vc{t}_{\mathcal{I}},\vc{x},\vc{\alpha}\right) := \PR\left(\vc{\tau} \in E_i \mid \vc{X} = \vc{x}, \vc{\tau}^{(\mathcal{I})} = \vc{t}_{\mathcal{I}}\right).
\]
The latter are simple and explicit for independent triggers:
\begin{equation}\label{eq:cond_prob_indep}
	\pi_{\mathcal{I}}\left(\vc{t}_{\mathcal{I}},\vc{x},\vc{\alpha}\right) = I_{\{\tau^{(i)}\leq T\ \forall i \in \mathcal{I}\}} \prod_{j \notin \mathcal{I}} \exp\left\{-\Lambda^{(j)}(T, \vc{X}, \aalpha)\right\},
\end{equation}
while in general one needs all conditional tail functions of the copula:
\[
	\pi_{\mathcal{I}}\left(\vc{t}_{\mathcal{I}},\vc{x},\vc{\alpha}\right) = I_{\{\tau^{(i)}\leq T\ \forall i \in \mathcal{I}\}}\bar{C}^{\mathcal{I}}_{\vc{X},\vc{\aalpha}}\left(u_{\mathcal{I}^c}(T,\vc{x},\vc{\alpha}) \mid u_{\mathcal{I}}(\vc{t}_{\mathcal{I}},\vc{x},\vc{\alpha}) \right),
\]	
with
\[
	 \bar{C}^{\mathcal{I}}_{\vc{X},\vc{\aalpha}}\left(u_{\mathcal{I}^c} \mid u_{\mathcal{I}} \right) := \PR_{\vc{U}\sim C_{\vc{X},\vc{\alpha}}}\bigg[U_j > u_j \ \forall j \notin \mathcal{I} \mid U_{i} = u_i \ \forall i \in \mathcal{I} \bigg].
\]
In this section we are assuming that $f$ depends on the exact value of $t_i$ only for those $i$ which are within maturity. In formulas, this means that functions $f_{\mathcal{I}}$ exists such that
\[
	f(\vc{t},\vc{x},\vc{\alpha}) = f_{\mathcal{I}}(\vc{t}_{\mathcal{I}},\vc{x},\vc{\alpha}) \qquad \forall \vc{t} \in E_{\mathcal{I}}.
\]
Then
\[
	p = \E\left\{ \sum_{\mathcal{I}\subseteq\{1,\dots,\bar{\imath}\}} f_{\mathcal{I}}\left(\vc{\tau}^{(\mathcal{I})}, \vc{X}, \aalpha\right) I_{\vc{\tau}\in E_{\mathcal{I}}} \right\} = \E\left\{ \sum_{\mathcal{I}\subseteq\{1,\dots,\bar{\imath}\}} \left[f_{\mathcal{I}} \pi_{\mathcal{I}}\right]\left(\vc{\tau}^{(\mathcal{I})},\vc{x},\vc{\alpha}\right) \right\},
\]
and differentiation by conditioning gives \eqref{eq:multiple} with $w(\vc{\tau},\vc{X},\aalpha)$ defined on $\{\vc{\tau}\in E_{\mathcal{I}}\}$ as
\begin{equation}\label{eq:log_density_censored_copula}
	\log \rho^{(\mathcal{I})}_{\vc{X},\aalpha}\left(\vc{u}(\vc{\tau}^{(\mathcal{I})},\vc{X},\vc{\alpha})\right)
	+ \sum_{i \in \mathcal{I}} \left[-\Lambda^{(i)} +\log\lambda^{(i)}\right](\tau^{(i)}, \vc{X}, \aalpha) + \log \pi_{\mathcal{I}}\left(\vc{\tau}^{(\mathcal{I})},\vc{x},\vc{\alpha}\right),
\end{equation}
where $\rho^{(\mathcal{I})}_{\vc{X},\aalpha}$ is the density of the restriction of the copula to the components in $\mathcal{I}$.

\begin{rem}[Independent triggers] 
Inserting \eqref{eq:cond_prob_indep} into \eqref{eq:log_density_censored_copula} one gets back \eqref{eq:log_density_censored_indep} without relying on the artificial model \eqref{eq:censored_intensity}.
\end{rem}

\begin{rem}[Zeroing terms]\label{rem:vanishing}
In estimator \eqref{eq:multiple}, $w$ appears multiplied by $f$. This implies that one does not need to compute \eqref{eq:log_density_censored_copula} for sets $\mathcal{I}$ such that $f_{\mathcal{I}}\equiv 0$.

Such zero-valued subsets can also be created artificially by a suitable splitting of the payoff. For instance, one can always write
\[
	p = \E[f_{\emptyset}\left(\vc{X}, \aalpha\right)] + \E[f\left(\vc{\tau},\vc{X}, \aalpha\right)-f_{\emptyset}\left(\vc{X}, \aalpha\right)] =: \E[f_{\emptyset}\left(\vc{X}, \aalpha\right)] + \E[f^{(0)}\left(\vc{\tau},\vc{X}, \aalpha\right)],
\]
where the first integrand does not depend on default times, while the second one is null on $\{\vc{\tau} \in E_{\emptyset}\}$ by construction.

More generally, if one has a payoff $f^{(k-1)}$ such that $f^{(k-1)}_{\mathcal{I}} \equiv 0$ for all $|\mathcal{I}| < k$, the split
\[
	f^{(k-1)}\left(\vc{\tau},\vc{X}, \aalpha\right) = \sum_{|\mathcal{I}|=k} f^{(k-1)}_{\mathcal{I}}\left(\vc{\tau}^{(\mathcal{I})},\vc{X}, \aalpha\right) I_{\{\vc{\tau}^{(i)}\leq T\ \forall i \in \mathcal{I}\}} + f^{(k)}\left(\vc{\tau},\vc{X}, \aalpha\right)
\]
defines a remainder $ f^{(k)}$ such that 
\[
	 f^{(k)}\left(\vc{\tau},\vc{X}, \aalpha\right) = f^{(k-1)}_{\mathcal{J}}\left(\vc{\tau}^{(\mathcal{J})},\vc{X}, \aalpha\right) - \sum_{|\mathcal{I}|=k,  \mathcal{I}\subseteq\mathcal{J}} f^{(k-1)}_{\mathcal{I}}\left(\vc{\tau}^{(\mathcal{I})},\vc{X}, \aalpha\right) \qquad \text{on $\{\vc{\tau}\in E_\mathcal{J}\}$},
\]
hence trivially $f^{(k)}_{\mathcal{J}}$ remains zero for $|\mathcal{J}| < k$ but is also zero for $|\mathcal{J}| = k$. This gives a recursive procedure to zero out as many terms as one wishes; which can be beneficial e.g.~if the conditional probabilities $\pi_{\mathcal{I}}$ are less tractable when $\mathcal{I}^c$ has high cardinality, as in \cref{ex:gaussian} below.

In fact, the resulting additive decomposition of the original payoff
\begin{equation}\label{eq:decomposition}
	f = \sum_{|\mathcal{I}| \leq k} f^{(|\mathcal{I}|-1)}_{\mathcal{I}}\left(\vc{\tau}^{(\mathcal{I})},\vc{X}, \aalpha\right) I_{\{\vc{\tau}^{(i)}\leq T\ \forall i \in \mathcal{I}\}} + f^{(k)}\left(\vc{\tau},\vc{X}, \aalpha\right)
\end{equation}
can be written down explicitly in terms of the $f_{\mathcal{I}}$. Indeed, we can prove that the generic addend is
\begin{equation}\label{eq:decomposition_addend}
	f^{(|\mathcal{I}|-1)}_{\mathcal{I}}\left(\vc{\tau}^{(\mathcal{I})},\vc{X}, \aalpha\right) = \sum_{\mathcal{J}\subseteq\mathcal{I}} (-1)^{|\mathcal{I}\setminus\mathcal{J}|}f_{\mathcal{J}}\left(\vc{\tau}^{(\mathcal{J})},\vc{X}, \aalpha\right)
\end{equation}
for all $\mathcal{I}$ with $|\mathcal{I}| = k$ by induction on $k$ as follows. In the base case $k=0$, the result is trivial because $f^{(-1)}=f$. For $|\mathcal{I}|=k+1$ one uses \eqref{eq:decomposition} to get
\[
	f^{(k)}_{\mathcal{I}}\left(\vc{\tau}^{(\mathcal{I})},\vc{X}, \aalpha\right) =
	f_{\mathcal{I}}\left(\vc{\tau}^{(\mathcal{I})},\vc{X}, \aalpha\right) - \sum_{\mathcal{J}\subset\mathcal{I}} f^{(|\mathcal{J}|-1)}_{\mathcal{J}}\left(\vc{\tau}^{(\mathcal{J})},\vc{X}, \aalpha\right),
\]
which by the induction hypothesis \eqref{eq:decomposition_addend} equals
\begin{multline*}
	f_{\mathcal{I}}\left(\vc{\tau}^{(\mathcal{I})},\vc{X}, \aalpha\right) - \sum_{\mathcal{J}\subset\mathcal{I}} \sum_{\mathcal{K}\subseteq\mathcal{J}} (-1)^{|\mathcal{J}\setminus\mathcal{K}|}f_{\mathcal{K}}\left(\vc{\tau}^{(\mathcal{K})},\vc{X}, \aalpha\right) \\
	= f_{\mathcal{I}}\left(\vc{\tau}^{(\mathcal{I})},\vc{X}, \aalpha\right) - \sum_{\mathcal{K}\subset\mathcal{I}} f_{\mathcal{K}}\left(\vc{\tau}^{(\mathcal{K})},\vc{X}, \aalpha\right) \sum_{\mathcal{K}\subseteq\mathcal{J}\subset\mathcal{I}} (-1)^{|\mathcal{J}\setminus\mathcal{K}|}:
\end{multline*}
this gives the conclusion by the elementary equality\footnote{Recall that for any set $\mathcal{A}$ one has $\sum_{\mathcal{B}\subseteq\mathcal{A}}(-1)^{|\mathcal{B}|}=0$, e.g.~by expansion of $(1-1)^{|\mathcal{A}|}=0$.}
\[
	 \sum_{\mathcal{K}\subseteq\mathcal{J}\subset\mathcal{I}} (-1)^{|\mathcal{J}\setminus\mathcal{K}|} = \sum_{\mathcal{H}\subset\mathcal{I}\setminus\mathcal{K}} (-1)^{|\mathcal{H}|} = \sum_{\mathcal{H}\subseteq\mathcal{I}\setminus\mathcal{K}} (-1)^{|\mathcal{H}|} - (-1)^{|\mathcal{I}\setminus\mathcal{K}|} = - (-1)^{|\mathcal{I}\setminus\mathcal{K}|}.
\]
\end{rem}

\begin{exmp}[Bilateral CVA]
In application \eqref{eq:bcva}, one already has $f_{\emptyset} = f_{\{2\}}=0$. Most used copulas have an analytical expression for $\smash{\bar{C}^{\{1\}}_{\vc{X},\vc{\aalpha}}\left(u_2 \mid u_1 \right)}$ (see e.g.~\cref{ex:gaussian}), so no manipulations are required. Anyway, zeroing also $f_{\{1\}}$ via \cref{rem:vanishing} would lead to a financially meaningful split
\begin{equation*}
	-\lgd\cdot\E\left[D(0,\tau^{(1)})(\NPV_{\tau^{(1)}})^{+}I_{\{\tau^{(1)} \leq T\}}\right] +\lgd\cdot \E\left[D(0,\tau^{(1)})(\NPV_{\tau^{(1)}})^{+}I_{\{\tau^{(2)}<\tau^{(1)} \leq T\}}\right]
\end{equation*}
of $p$ as the sum of unilateral CVA \eqref{eq:ucva} and a \quoted{DVA of CVA} correction.
\end{exmp}

\begin{exmp}[Gaussian copula]\label{ex:gaussian}
For illustration and practical convenience, we write down explicitly \eqref{eq:log_density_censored_copula} for the most popular copula in credit modelling, which is the unique copula of a standard normal vector $\mathcal{N}(\vc{0},\mt{R})$ with full rank correlation matrix $\mt{R}$.

It is trivial that the copula of a sub-vector $\vc{u}_{\mathcal{I}}$ is Gaussian with correlation matrix $\mt{R}_{\mathcal{I}\mathcal{I}}$ for all $\mathcal{I}$. The density of the Gaussian copula is well known:
\[
	\log \rho^{(\mathcal{I})}_{\vc{X},\aalpha}(\vc{u}_{\mathcal{I}}) = -\frac12\left[\tr{\Phi^{-1}(\vc{u}_{\mathcal{I}})}\left(\mt{R}_{\mathcal{I}\mathcal{I}}^{-1}-\mathrm{Id}\right)\Phi^{-1}(\vc{u}_{\mathcal{I}})+\log\mathrm{det}\left(\mt{R}_{\mathcal{I}\mathcal{I}}\right)\right]
\]
where the standard normal quantile function $\Phi^{-1}$ is applied componentwise.

The required conditional probabilities are as follow:
\begin{align*}
\bar{C}^{\mathcal{I}}_{\vc{X},\vc{\aalpha}}\left(u_{\mathcal{I}^c} \mid u_{\mathcal{I}} \right) &= \PR_{\vc{X}\sim \mathcal{N}(\vc{0},\mt{R})}\bigg[X_j > \Phi^{-1}(u_j) \ \forall j \notin \mathcal{I} \mid X_{i} = \Phi^{-1}(u_i) \ \forall i \in \mathcal{I} \bigg]\\
&= \PR_{\vc{X}_{\mathcal{I}^c}\sim \mathcal{N}\left(\mt{R}_{\mathcal{I}^c\mathcal{I}}\mt{R}_{\mathcal{I}\mathcal{I}}^{-1}\Phi^{-1}(\vc{u}_{\mathcal{I}}),\mt{R}_{\mathcal{I}^c\mathcal{I}^c}-\mt{R}_{\mathcal{I}^c\mathcal{I}}\mt{R}_{\mathcal{I}\mathcal{I}}^{-1}\mt{R}_{\mathcal{I}\mathcal{I}^c}\right)}\bigg[\vc{X}_{\mathcal{I}^c} > \Phi^{-1}\left(\vc{u}_{\mathcal{I}^c}\right) \bigg]\\
&= \PR_{\vc{Y}_{\mathcal{I}^c}\sim \mathcal{N}\left(\vc{0},\mt{R}_{\mathcal{I}^c\mathcal{I}^c}-\mt{R}_{\mathcal{I}^c\mathcal{I}}\mt{R}_{\mathcal{I}\mathcal{I}}^{-1}\mt{R}_{\mathcal{I}\mathcal{I}^c}\right)}\bigg[\vc{Y}_{\mathcal{I}^c} \leq \mt{R}_{\mathcal{I}^c\mathcal{I}}\mt{R}_{\mathcal{I}\mathcal{I}}^{-1}\Phi^{-1}(\vc{u}_{\mathcal{I}}) - \Phi^{-1}\left(\vc{u}_{\mathcal{I}^c}\right) \bigg].
\end{align*}
These normal cumulative distribution functions are not available in closed form for $|\mathcal{I}^{c}| > 1$, but all of them are numerically tractable at least for $|\mathcal{I}^{c}| \leq 3$ \citep{Genz2004numerical}.
So one is safe if $\bar{\imath}\leq 3$, while with four or more underlying names one may consider the procedure in \cref{rem:vanishing} to zero out $f_{\mathcal{I}}$ for $|\mathcal{I}|<\bar{\imath}-3$.
\end{exmp}

\section{Second order model sensitivities}\label{sec:gamma}

In this section we focus on the estimation of $\iderXX{p}{\vc{\alpha}}$. To derive efficient estimators, we add the following hypothesis which is very often satisfied in practice:
\begin{hp}
The set of parameters $\vc{\alpha}$ is the union of two vectors $\vc{\theta}$ and $\vc{\psi}$ such that:
\begin{enumerate}
\item $\ider{f}{\vc{\theta}} \equiv 0$: the payoff has no direct dependence on $\vc{\theta}$;
\item $\ider{\lambda^{(i)}}{\vc{\psi}} \equiv 0$ for all $i$: default intensities do not depend on $\vc{\psi}$.
\end{enumerate}
\end{hp}
The main motivating example is the CVA of a portfolio which does not include credit-linked products.

Specializing the results in \cref{sec:multiple,sec:censored}, we note that the first derivative with respect to $\vc{\psi}$ will have only the functional dependence term, while the one with respect to $\vc{\theta}$ will have only the conditional log-likelihood term:
\begin{equation*}
	\der{p}{\vc{\psi}} = \E\left\{\der{f}{\vc{\psi}}(\vc{\tau},\vc{X},\aalpha)\right\},\qquad
	\der{p}{\vc{\theta}} = \E\left\{f(\vc{\tau},\vc{X},\aalpha) \der{w}{\vc{\theta}}(\vc{\tau},\vc{X},\aalpha)\right\}.
\end{equation*}
These are again expected values of functions of $(\vc{\tau},\vc{X},\vc{\alpha})$, so we can apply our conditional differentiation results once more and get\footnote{We adopt a maybe counter-intuitive convention which is common to ease algorithmic differentiation, according to which entry $(i,j)$ of $\iderXY{f}{\vc{x}}{\vc{y}}$ is $\iderXY{f}{x_j}{y_i}$.}
\begin{gather}
	\derXY{p}{\vc{\theta}}{\vc{\psi}} = \E\left[\tr{\left(\der{f}{\vc{\psi}}(\vc{\tau},\vc{X},\aalpha)\right)\!} \der{w}{\vc{\theta}}(\vc{\tau},\vc{X},\aalpha) \right],\label{eq:mixed_gamma}\\
	\derXX{p}{\vc{\theta}} = \E\left\{f(\vc{\tau},\vc{X},\aalpha) \left[ \derXX{w}{\vc{\theta}} + \tr{\left(\der{w}{\vc{\theta}}\right)\!} \der{w}{\vc{\theta}} \right](\vc{\tau},\vc{X},\aalpha) \right\}\label{eq:pure_gamma}.
\end{gather}
The purely functional second order differentiation with respect to $\vc{\psi}$ is covered by the general analysis of Monte Carlo Hessians in \citet{Daluiso2020secondOrder}, and is therefore not studied in this paper.

\begin{rem}[Algorithmic complexity]
In the above formulas, the payoff $f$ is differentiated only once, which can be done efficiently by AAD. For the same reason, the gradients of $w$ in \eqref{eq:mixed_gamma}-\eqref{eq:pure_gamma} are of no concern.
The only problematic term could be the second derivative $\iderXX{w}{\vc{\theta}}$ in \eqref{eq:pure_gamma}; fortunately, the cost of $w$ is very often negligible when compared to the computational time of $f$: e.g.~in CVA, the latter includes the simulation of risk drivers and the evaluation of the portfolio.
Moreover, Cross Gammas \eqref{eq:mixed_gamma} are empirically more relevant than pure Credit Gammas \eqref{eq:pure_gamma} for typical CVA desks; see \cref{sec:results_2} for an empirical confirmation.
\end{rem}

\section{Market sensitivities}\label{sec:conversion}

In practice, pricing inputs $\aalpha$ are a function $\aalpha(\vc{m})$ of quotes $\vc{m}$;
the above algorithms give the derivatives of $p$ with respect to $\aalpha$, while the quantities of interest would be the derivatives of the pricing function inclusive of calibration:
\[
	P(\vc{m}) := p(\aalpha(\vc{m})).
\]

Of course, if the map $\vc{m}\to\aalpha$ is a plain sequence of elementary differentiable operations, one has AAD to convert sensitivities of $p$ into sensitivities of $P$. But this is seldom the case: almost always, that map involves iterative procedures to fit the price of a set of market instruments. Therefore, in this section we discuss other ways to compute the gradient and Hessian of $P$ from those of $p$.

\subsection{First order}\label{sec:conversion_1}

In this subsection we present two algorithms for first order sensitivities.

The most well known conversion rule covers parameters which are chosen to reprice perfectly a set of market instruments: for instance, yield curves bootstrapped from a set of swap rates, or deterministic intensity models bootstrapped from a term structure of Credit Default Swaps. Indeed in such case $\aalpha(\vc{m})$ is characterized by a set of equations $\vc{b}(\aalpha,\vc{m}) = \zzero$, and one can apply the implicit function theorem \citep{Henrard2011ADimplicitFunction}:
\begin{equation}\label{eq:implicit_1}
		\der{P}{\vc{m}} = \der{p}{\aalpha} \cdot \der{\aalpha}{\vc{m}} = - \der{p}{\aalpha}\left(\der{\vc{b}}{\aalpha}\right)^{-1}\der{\vc{b}}{\vc{m}}.
\end{equation}
Note that $\vc{b}$ has usually a fast analytical implementation, as it must have been called several times in the numerical search of the root $\aalpha$: this means that the Jacobian of $\vc{b}$ appearing in the above formula can be computed efficiently.

Less standard generalizations to best fit calibration exist \citep{Henrard2013ADleastSquare, Daluiso2016modelToMarket}, but we will not attempt a second order generalization of them in the present paper. This is because for our main application, a pragmatic finite difference conversion is enough: namely, for each direction $\vc{\mu}$ of interest in the space of quotes $\vc{m}$, one can take $h\in\R$ small and compute:
\begin{equation}\label{eq:fd_1}
	\der{P}{\vc{m}} = \der{p}{\aalpha} \cdot \der{\aalpha}{\vc{m}} \cdot \vc{\mu} \approx \der{p}{\aalpha} \cdot \frac{\aalpha(\vc{m}+h\vc{\mu}) - \aalpha(\vc{m})}{h},
\end{equation}
where equality holds in the limit $h\to 0$.

This is acceptable when the implementation of the map $\vc{m}\to\aalpha$ is much faster than $p$, which is for sure the case if $p$ is CVA pricing. Moreover, the typical motivation for best fit calibration is parsimony: hence the number of parameters to which \eqref{eq:implicit_1} cannot be applied should not be very large in real applications.

\subsection{Second order}\label{sec:conversion_2}

From $\aalpha = \aalpha(\vc{m})$, applying twice the chain rule:\footnote{We choose to keep a synthetic index-less notation, at the cost of a slight ambiguity on the contraction coordinate in some products of tensors or rank 3. The meaning should be clear from the contest; anyway, we try to ease the reading by reserving the dot notation for multiplications contracting the \quoted{numerator} of the second derivative of a vector valued function.}

\begin{equation}\label{eq:chain_rule_2}
	\derXX{P}{\vc{m}} = \der{p}{\aalpha}\cdot \derXX{\aalpha}{\vc{m}} + \tr{\left(\der{\aalpha}{\vc{m}}\right)\!}\derXX{p}{\aalpha}\left(\der{\aalpha}{\vc{m}}\right).
\end{equation}

We first cover the simpler finite difference implementation. For each direction $\vc{\mu}$ of interest in the space of quotes $\vc{m}$, one can take $h\in\R$ small and compute:
\begin{align}
	\derXX{P}{\vc{m}}h\vc{\mu} &= \der{p}{\aalpha}\cdot \derXX{\aalpha}{\vc{m}}h\vc{\mu} + \tr{\left(\der{\aalpha}{\vc{m}}\right)\!}\derXX{p}{\aalpha}\left(\der{\aalpha}{\vc{m}}\right)h\vc{\mu}\nonumber \\
	& \approx \der{p}{\aalpha}\cdot \left[ \der{\aalpha}{\vc{m}}(\vc{m}+h\vc{\mu}) - \der{\aalpha}{\vc{m}}(\vc{m}) \right] + \tr{\left(\der{\aalpha}{\vc{m}}\right)\!} \derXX{p}{\aalpha} \left[ \aalpha(\vc{m}+h\vc{\mu}) - \aalpha(\vc{m}) \right],\label{eq:1}
\end{align}
where equality holds to first order in $h$.
Now note that the propagation of first order sensitivities is the map
\[
	\bar{\aalpha}(p_{\aalpha},\vc{m}) := p_{\aalpha} \cdot \der{\aalpha}{\vc{m}}(\vc{m});
\]		
hence if one has an implementation of $\bar{\aalpha}$, say \eqref{eq:implicit_1}, then \eqref{eq:1} is readily available as
\begin{equation}\label{eq:fd_2}
	\bar{\aalpha}\left(\derXX{p}{\aalpha} \left[\aalpha(\vc{m}+h\vc{\mu}) - \aalpha(\vc{m}) \right]-\der{p}{\aalpha},\vc{m}\right) + \bar{\aalpha}\left(\der{p}{\aalpha},\vc{m}+h\vc{\mu}\right).
\end{equation}

Alternatively, analytical formulas can be derived if all calibration steps are perfect fits. For concreteness, we suppose that $\vc{m}$ is partitioned as $(\vc{c}, \vc{q})$, where $\vc{c}$ represents credit quotes, analogously to the partitioning $\aalpha = (\vc{\theta},\vc{\psi})$ of model parameters; in particular, we assume that $\vc{\psi}$ does not depend on $\vc{c}$. As in the rest of the paper we concentrate on second order sensitivities involving $\vc{c}$ at least once, i.e.~$\iderXX{P}{\vc{c}}$ and ~$\iderXY{P}{\vc{q}}{\vc{c}}$. Specializing \eqref{eq:chain_rule_2} we obtain:
\begin{align}
\label{eq:mkt_gamma}\derXX{P}{\vc{c}} &= \der{p}{\vc{\theta}} \cdot \derXX{\vc{\theta}}{\vc{c}} + \tr{\left(\der{\vc{\theta}}{\vc{c}}\right)\!}\derXX{p}{\vc{\theta}}\left(\der{\vc{\theta}}{\vc{c}}\right),\\
\label{eq:mkt_cross_gamma}\derXY{P}{\vc{q}}{\vc{c}} &= \der{p}{\vc{\theta}}\cdot \derXY{\vc{\theta}}{\vc{q}}{\vc{c}} + \tr{\left(\der{\vc{\theta}}{\vc{c}}\right)\!}\derXX{p}{\vc{\theta}}\left(\der{\vc{\theta}}{\vc{q}}\right) + \tr{\left(\der{\vc{\theta}}{\vc{c}}\right)\!}\derXY{p}{\vc{\psi}}{\vc{\theta}}\left(\der{\vc{\psi}}{\vc{q}}\right).
\end{align}
The problematic addend in both equations is the first one, as it involves the Hessian of the calibration of $\vc{\theta}$, while the other addends are readily available given the model Hessians computed in the previous sections. Indeed, these Hessians appear multiplied by sub-matrices of $\ider{\vc{\alpha}}{\vc{m}}$, and as above these multiplications can be obtained by calling repeatedly the first order propagation routines.

The addends depending on calibration Hessians can be expressed as
\[
	\der{P_c}{\vc{c}}\left(\vc{c},\vc{q},\der{p}{\vc{\theta}}\right) \qquad \text{and} \qquad \der{P_c}{\vc{q}}\left(\vc{c},\vc{q},\der{p}{\vc{\theta}}\right)
\]
where in the below definition $p_{\vc{\theta}}$ is interpreted as a further argument not depending on $\vc{c},\vc{q}$:
\[
	P_c\left(\vc{c},\vc{q},p_{\vc{\theta}}\right) := \tr{\left(p_{\vc{\theta}}\der{\vc{\theta}}{\vc{c}}\right)}.
\]

Let us suppose that $\vc{\theta}$ is obtained from $\vc{c}$ by a perfect fit:
\[
	\vc{b}(\vc{c}, \vc{q}, \vc{\theta}(\vc{c},\vc{q})) = \vc{0}.
\]
Then one can remember the expression for the differential of the function $\mt{s}(\mt{A},\mt{M}) := \mt{M}^{-1} \mt{A}$ applied to the generic pair $[\delta\mt{A},\delta\mt{M}]$:
\[
	\diff{}\mt{s}\left[\delta\mt{A},\delta\mt{M}\right] = \mt{M}^{-1}\delta\mt{A} - \mt{M}^{-1}(\delta\mt{M})\mt{s},
\]
to differentiate the relation
\[
	\tr{P}_{\vc{c}} = -p_{\vc{\theta}}\left(\der{\vc{b}}{\vc{\theta}}\right)^{-1}\der{\vc{b}}{\vc{c}} = -p_{\vc{\theta}}\, \mt{s}\left(\der{\vc{b}}{\vc{c}},\der{\vc{b}}{\vc{\theta}}\right).
\]
In particular, for every $\delta{\vc{c}}$
\begin{align*}
	\diff_c \tr{P}_{\vc{c}}[\delta\vc{c}] = &-p_{\vc{\theta}}\left(\der{\vc{b}}{\vc{\theta}}\right)^{-1} \cdot \left[ \derXY{\vc{b}}{\vc{\theta}}{\vc{c}} \der{\vc{\theta}}{\vc{c}} + \derXX{\vc{b}}{\vc{c}} \right]\delta\vc{c}\\
	&+ p_{\vc{\theta}} \left(\der{\vc{b}}{\vc{\theta}}\right)^{-1} \cdot \left[ \derXX{\vc{b}}{\vc{\theta}} \der{\vc{\theta}}{\vc{c}} + \derXY{\vc{b}}{\vc{c}}{\vc{\theta}} \right]\delta\vc{c}\left(-\der{\vc{\theta}}{\vc{c}}\right).
\end{align*}
The $j$-th column of $\ider{P_c}{\vc{c}}$ is now obtained by using the above relation for $\delta\vc{c}$ equal to the $j$-th element of the canonical basis (and transposing the resulting row vector):
\begin{equation}\label{eq:mkt_gamma_addend}
	\der{\tr{P}_{\vc{c}}}{c_j} = -p_{\vc{\theta}}\left(\der{\vc{b}}{\vc{\theta}}\right)^{-1} \cdot \left\{ \left[ \derXY{\vc{b}}{\vc{\theta}}{\vc{c}} \der{\vc{\theta}}{c_j} + \derXY{\vc{b}}{c_j}{\vc{c}} \right] + \left[ \derXX{\vc{b}}{\vc{\theta}} \der{\vc{\theta}}{c_j} + \derXY{\vc{b}}{c_j}{\vc{\theta}} \right]\left(\der{\vc{\theta}}{\vc{c}}\right)\right\},
\end{equation}
which therefore is the contribution of the first addend of \eqref{eq:mkt_gamma} to the Hessian of $P$ with respect to $\vc{c}$, once evaluated in $p_{\vc{\theta}} = \ider{p}{\vc{\theta}}$. A completely analogous computation gives
\begin{align*}
	\diff_{\vc{q}} \tr{P}_{\vc{c}}[\delta\vc{q}] = &-p_{\vc{\theta}}\left(\der{\vc{b}}{\vc{\theta}}\right)^{-1} \cdot \left[\derXY{\vc{b}}{\vc{\theta}}{\vc{c}} \der{\vc{\theta}}{\vc{q}} + \derXY{\vc{b}}{\vc{q}}{\vc{c}} \right] \delta\vc{q}\\
	&+ p_{\vc{\theta}} \left(\der{\vc{b}}{\vc{\theta}}\right)^{-1} \cdot \left[\derXX{\vc{b}}{\vc{\theta}}\der{\vc{\theta}}{\vc{q}} + \derXY{\vc{b}}{\vc{q}}{\vc{\theta}} \right] \delta\vc{q}\left(-\der{\vc{\theta}}{\vc{c}}\right),
\end{align*}
and hence the $j$-th column of the first addend of \eqref{eq:mkt_cross_gamma}, again to be evaluated in $p_{\vc{\theta}} = \ider{p}{\vc{\theta}}$ and then transposed:
\begin{equation}\label{eq:mkt_cross_gamma_addend}
	-p_{\vc{\theta}}\left(\der{\vc{b}}{\vc{\theta}}\right)^{-1} \cdot \left\{\left[\derXY{\vc{b}}{\vc{\theta}}{\vc{c}} \der{\vc{\theta}}{q_j} + \derXY{\vc{b}}{q_j}{\vc{c}} \right] + \left[\derXX{\vc{b}}{\vc{\theta}}\der{\vc{\theta}}{q_j} + \derXY{\vc{b}}{q_j}{\vc{\theta}} \right]\left(\der{\vc{\theta}}{\vc{c}}\right) \right\}.
\end{equation}

\begin{rem}[Composition of calibration steps]
In the above formula, no second order differentiation with respect to $\vc{q}$ appears. This means that in practice if $\vc{b}(\vc{c}, \vc{q}, \vc{\theta})$ is implemented as $\vc{b}(\vc{c}, \vc{\psi}(\vc{q}), \vc{\theta})$, one can readily substitute everywhere
\[
	\der{}{\vc{q}} \rightarrow \der{}{\vc{\psi}}\der{\vc{\psi}}{\vc{q}}
\]
and the result still holds.
Once more, the multiplication by $\ider{\vc{\psi}}{\vc{q}}$ of any matrix is just a matter of applying first order sensitivity propagation to each of its rows.
\end{rem}

\section{Numerical tests}\label{sec:numerics}

In this section we present numerical evidence on the efficiency of estimators from \cref{sec:delta,sec:gamma}. Indeed, as they are based on Monte Carlo, one must check that their promising low cost-per-path is not offset by the need of a large number of paths to get a given accuracy, due to high statistical noise.

\subsection{Experimental setting}\label{sec:numerics_setting}

We consider a prototypical unilateral CVA pricing exercise specified as follows:
\begin{itemize}
	\item The netting set consists of a single At-The-Money (ATM) Overnight Indexed Swap with a notional of 100M EUR and a maturity of 10 years, receiving a fixed rate of 0.947\% and paying ESTR compounded annually.
	\item Interest rates are modelled by a one factor Hull\&White short rate model \citep{HullWhite1990pricingIRderivatives}, with mean reversion speed (0.0744) and volatility (0.0125) calibrated to few ATM swaptions. We assume that the zero rate $\log D(0,\cdot)$ is a piecewise linear function with levels $(\bar{r}_i)_i$ at a set of pillar dates $(T_i^r)_i$.
	\item Counterparty defaults obey piecewise constant deterministic hazard rates calibrated to Credit Default Swap quotes of an industrial counterparty with rating Ba and $\lgd = 60\%$. In analogy with rates, we parametrize the credit curve by the values $(\bar{\lambda}_j)_j$ of the \quoted{zero intensity} $\log \PR(\tau > \cdot)$ at a set of pillar dates $(T_j^{\lambda})_j$.
\end{itemize}
For reproducibility we provide the values of $T_i^r$, $T_j^\lambda$, $\bar{r}_i$ and $\bar{\lambda}_j$ in \cref{tab:zeros}.

\begin{table}
\centering \small
\begin{tabular}{c|c|c}%
    Label & Time ($T_i^r$) & Zero Rate ($\bar{r}_i$) %  table head
    \csvreader[head to column names,range=-38]{ESTR.csv}{}% use head of csv as column names, range up to 10Y
    {\\\pillar & \time & \zero}% specify your columns here
\end{tabular}\quad
\begin{tabular}{c|c|c}%
    Label & Time ($T_j^\lambda$) & Zero Intensity ($\bar{\lambda}_j$) %  table head
    \csvreader[head to column names,range=-7]{INDUSTRIAL_Ba.csv}{}% use head of csv as column names, range up to 10Y
    {\\\pillar & \time & \zero}% specify your columns here
\end{tabular}
\caption{Zero rate and zero intensity pillars.}\label{tab:zeros}
\end{table}

All Monte Carlo runs will consist of 100k scenarios, and uncertainties will be expressed as half confidence intervals at 98\% level, estimated via the central limit theorem from sample standard deviations. The performances will be always expressed as the product of computational time and Monte Carlo variance, as this is a measure which does not depend on the number of paths: one can interpret it as the time needed to get unit uncertainty. All computational times are obtained with a single-core Matlab prototype to make the measure more readable, although of course production implementations can easily split Monte Carlo paths across different threads or machines.

We compute sensitivities to both ESTR zero rates and counterparty \quoted{continuous par CDS spreads}, defined as $\bar{\lambda}_j\lgd$. We compare the following first order estimators:
\begin{enumerate}
	\item FD: lateral finite differences (i.e.~for each differentiation variable, bump its value up and reprice to compute an incremental ratio). % (1bp or 10bp on credit).
	\item CD: central finite differences (i.e.~for each differentiation variable, bump the value both up and down and reprice, to compute an incremental ratio which is more accurate even for larger bumps). %(1bp or 10bp on credit).
	\item AD: algorithmic differentiation + conditioning weights, i.e.~the algorithm described in \cref{sec:conditional}.
\end{enumerate}
Note that we did not implement \cref{sec:distributional}, which is of less practical interest because of \cref{rem:distributional_complexity}. The benchmarks FD and CD may be considered naive, but they represent the industry standard.
As for second order estimators, we test the following alternatives:
\begin{enumerate}
	\item FDAD: lateral finite differences on AD results.
	\item CDAD: central finite differences on AD results.
	\item AD2: second order algorithmic differentiation + conditioning weights, i.e.~the algorithm described in \cref{sec:gamma}.
\end{enumerate}
Here the benchmarks FDAD and CDAD are already quite sophisticated, as they rely on first order AAD (with conditioning for credit) to cut on computational times.

\subsection{Zero-th order result}\label{sec:results_0}

The base CVA value equals -535,594.26 EUR $\pm$ 14,402.64 EUR.

\subsection{First order results}\label{sec:results_1}

In this section we consider first order estimators; those based on finite differences are tried both with a small displacement of 1 basis point (bp, i.e.~$10^{-4}$), and with a larger one of 10 bp (i.e.~$10^{-3}$). This is because it is well known that for discontinuous payoffs, the bias induced by a larger bump can be more than compensated by lower Monte Carlo noise.

\Cref{fig:delta_results} shows the results. Rate sensitivities and their uncertainties look essentially undistinguishable among different methods: this is no surprise given the differentiable dependence of the payoff on zero rates, as finite differences in such case are close to the path-wise derivative, which is exactly what AD computes. On the other hand, credit sensitivities show roughly the same shape across estimators and are in fact compatible within Monte Carlo error, but their statistical uncertainties vary significantly: in particular, 1 bp finite differences are very noisy, while AD is by far more stable even than the biased 10 bp FD and CD.

\def\figwidth{\textwidth}

\begin{figure}
	\centering
	\includegraphics[width=\figwidth]{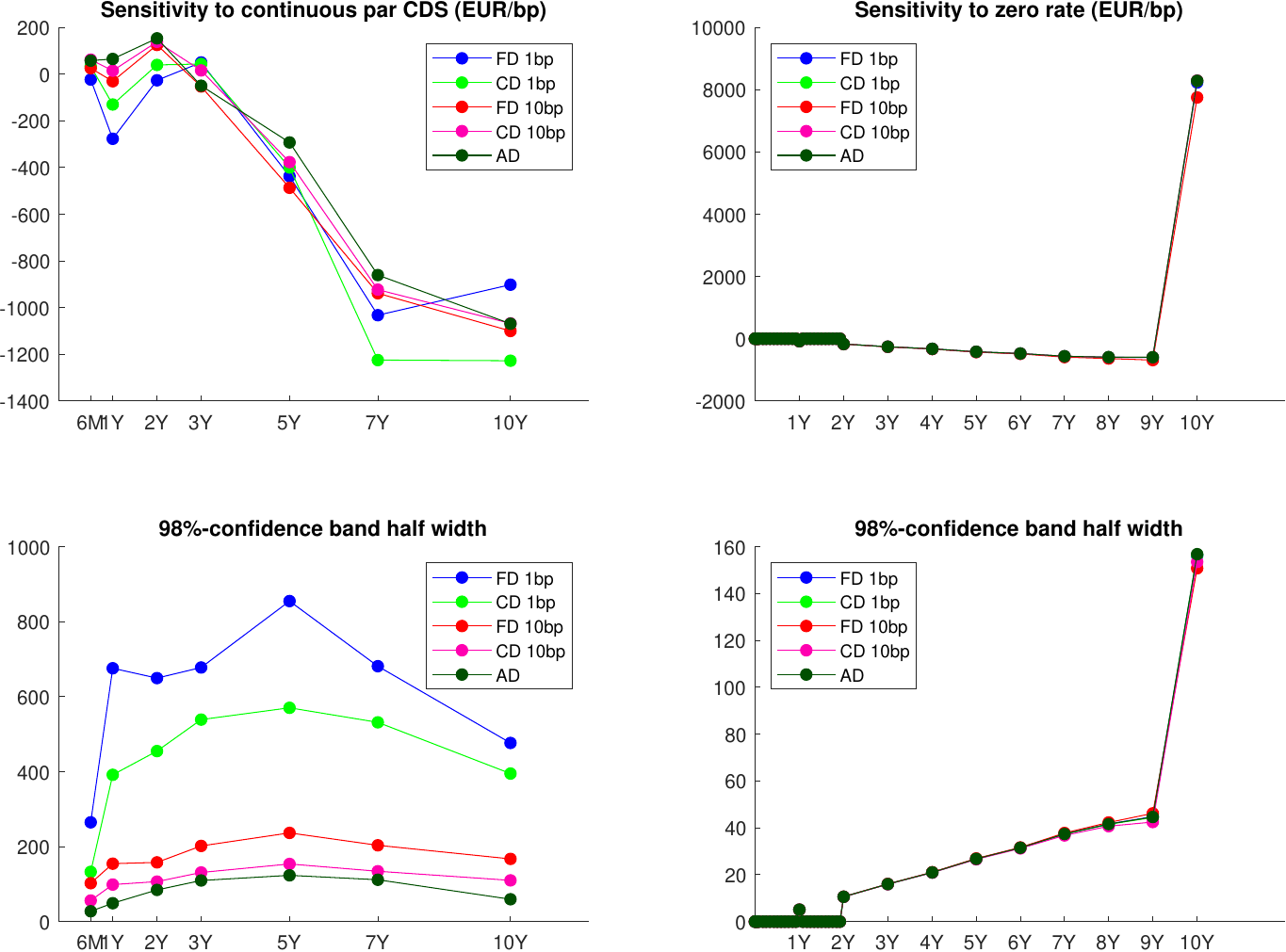}
	\caption{First order results. Graphs on the left refer to credit spread sensitivities, while graphs on the right refer to zero rate sensitivities. Top plots display average values, bottom plots display 98\%-confidence uncertainties over 100k paths.}\label{fig:delta_results}
\end{figure}

\begin{figure}
	\centering
	\includegraphics[width=\figwidth]{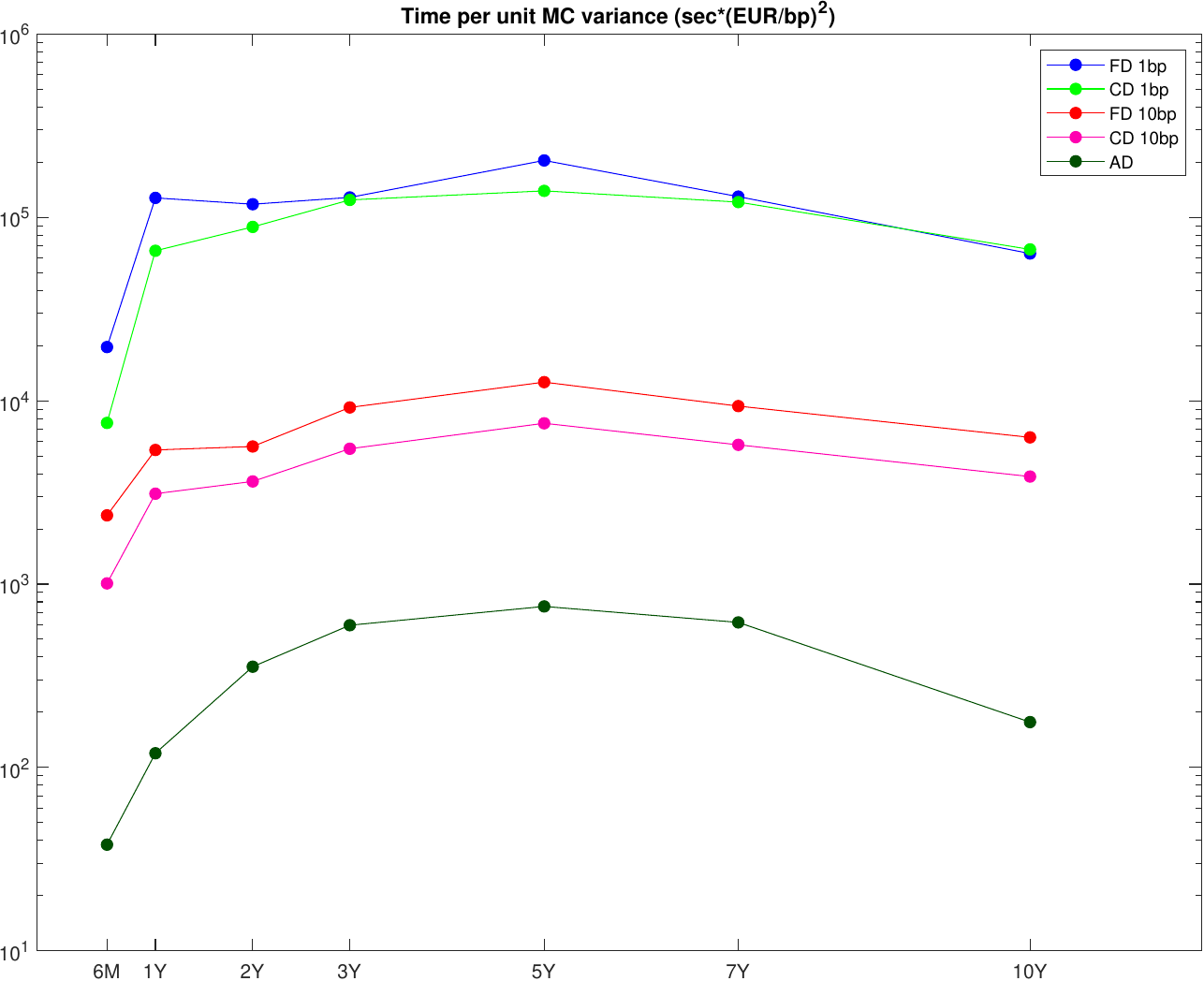}
	\caption{Credit Delta efficiency, defined as the multiplication of the Monte Carlo variance of a credit spread sensitivity by the total computational time needed to compute the full credit gradient. Note the log scale on the y axis.}\label{fig:delta_efficiency}
\end{figure}

These improvements in uncertainty look even more impressive when considering that they are accompanied by a significant performance gain: \cref{fig:delta_efficiency} multiplies the elapsed time to compute the full set of credit sensitivities by the Monte Carlo variance of each of them. The resulting normalized times  are plotted in logarithmic units: as one can see, for fixed desired uncertainty, AD is more than ten times faster than the best performing finite difference method.

\subsection{Second order results}\label{sec:results_2}

In this section we move to second order estimators; finite differences will be presented only for a displacement of 10 bp, as the results with 1 bp were already suboptimal for the gradient, and are much worse for the Hessian.

Since the computed matrices have many entries, we had to choose what to display. On the one hand, to give an indication of the overall results, some plots will sum derivatives across rows and/or columns: financially, these sums represent sensitivities where the movement of one or both the differentiation variables is assumed equal on all pillars (\quoted{parallel} sensitivity). On the other hand, as a sample of the most disaggregated results, some plots will concentrate on a single row and/or column; in this case we will chose the index of the \quoted{most important} credit or rate pillar, i.e.~that with largest absolute first order sensitivity.

\begin{figure}
	\centering
	\includegraphics[width=\figwidth]{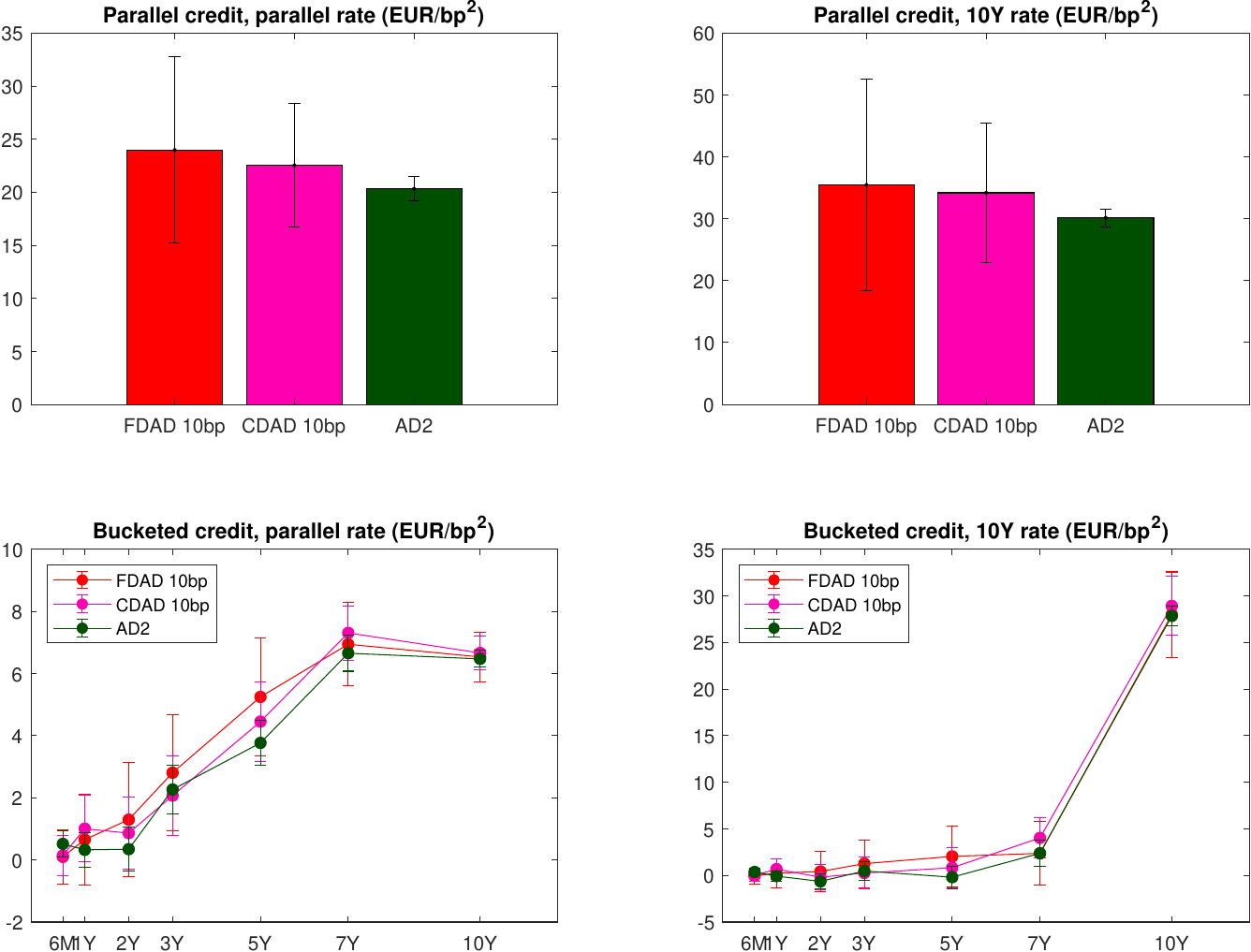}
	\caption{Second order mixed credit-rate sensitivities. Graphs on the left aggregate derivatives along the rate direction, while graphs on the right focus on the 10Y rate pillar. Top plots display values aggregated along the credit direction, bottom plots display disaggregated results. Error bars are 98\%-confidence bands over 100k paths.}\label{fig:cross_results}
\end{figure}

\begin{figure}
	\includegraphics[width=\figwidth]{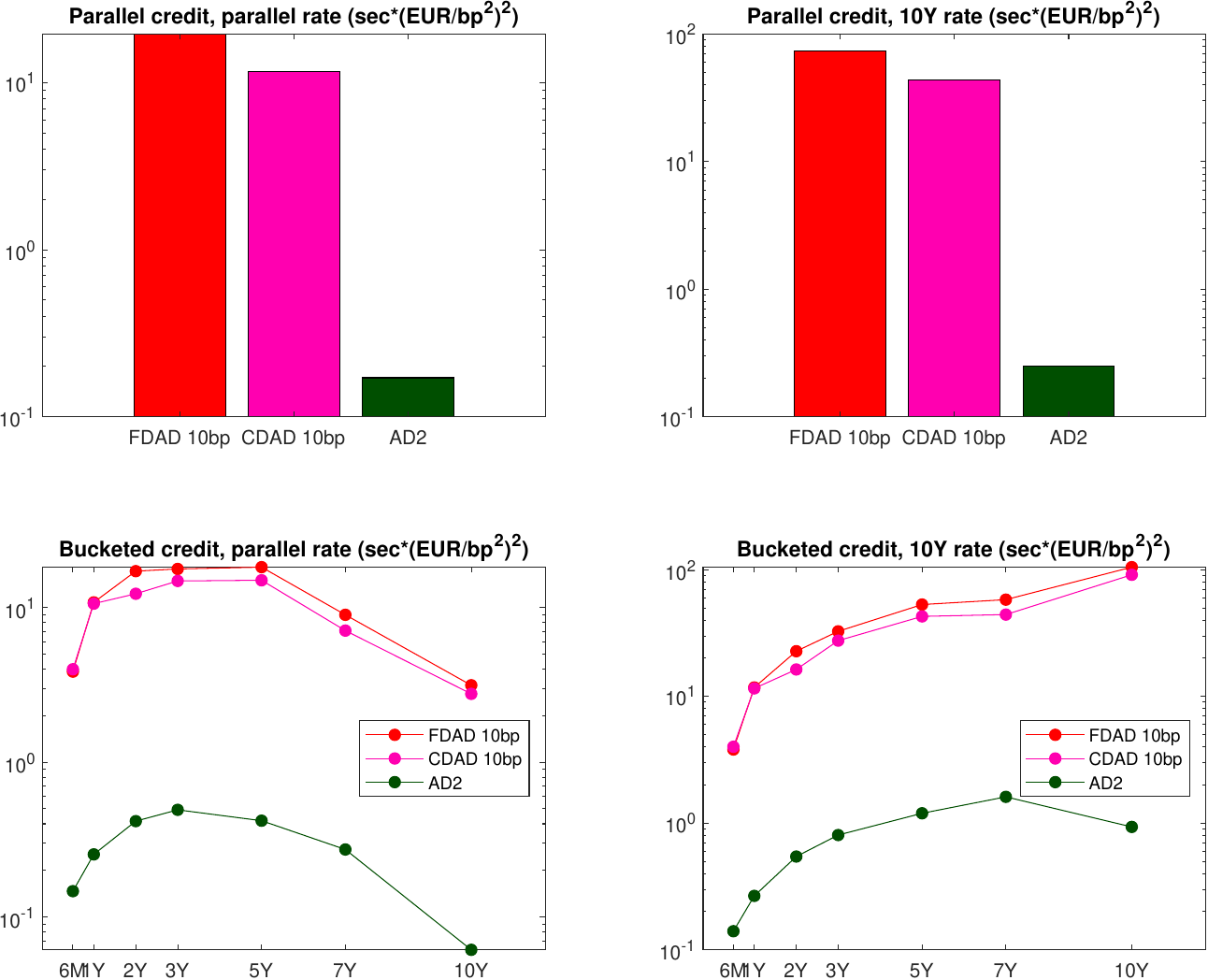}
	\caption{Cross Gamma efficiency, defined as the multiplication of the Monte Carlo variance of each sensitivity by the total computational time needed to compute the full credit-rates Hessian submatrix. Graphs on the left refer to the aggregated derivatives along the rate direction, while graphs on the right focus on the 10Y rate pillar. Top plots refer to sensitivities aggregated along the credit direction, bottom plots to disaggregated results. Note the log scale on the y axis.}\label{fig:cross_efficiency}
\end{figure}

\Cref{fig:cross_results} plots mixed derivatives where one differentiation variable is an interest rate and the other one is a credit spread.
Results are qualitatively similar, but the confidence bands of AD2 are much smaller than those of finite differences.
The corresponding semi-logarithmic efficiency plot in \cref{fig:cross_efficiency} shows improvements of at least two orders of magnitude.

\begin{figure}
	\centering
	\includegraphics[width=\figwidth]{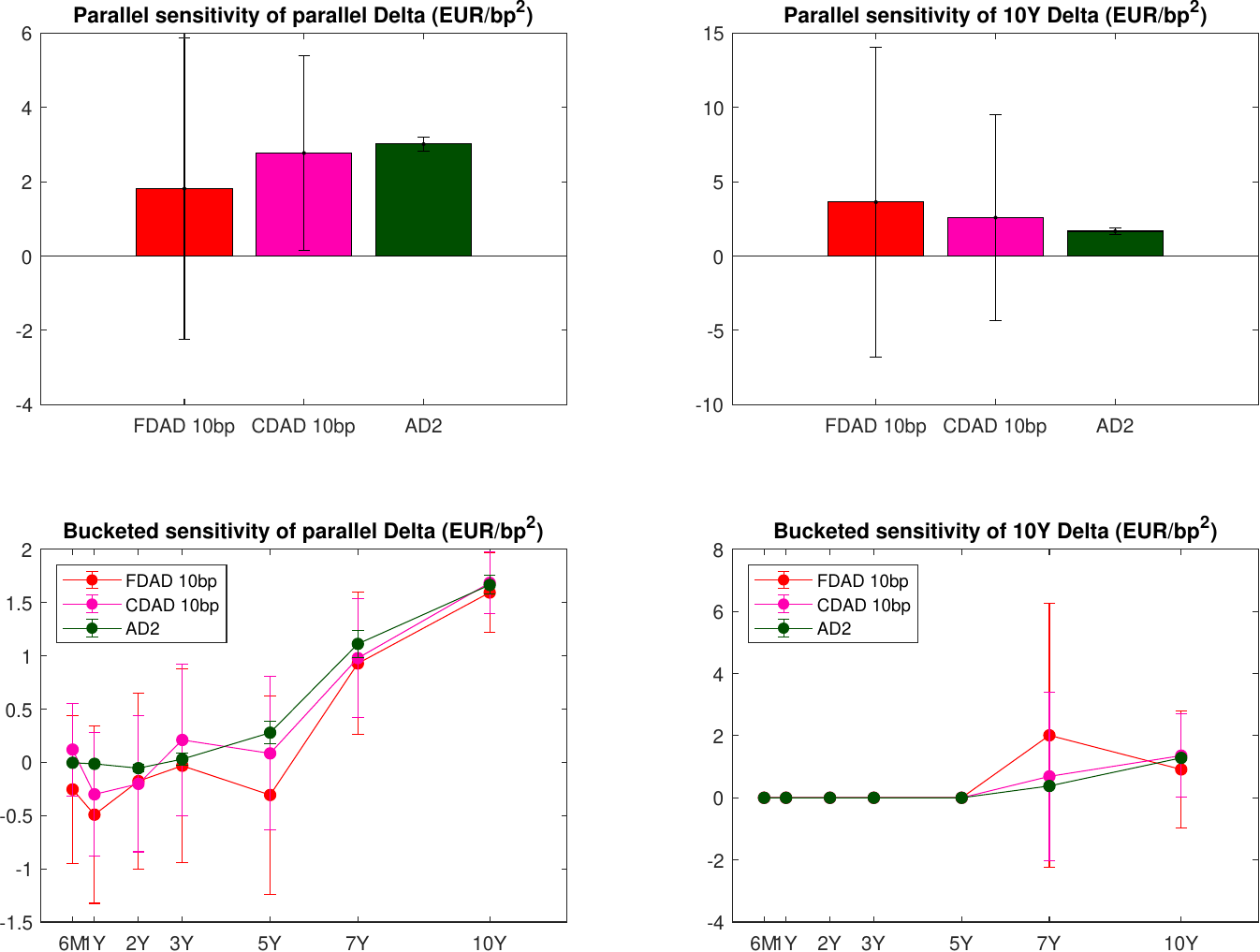}
	\caption{Second order credit sensitivities. Graphs on the left aggregate derivatives along the first differentiation variable, while graphs on the right focus on the 10Y credit pillar. Top plots display values aggregated along the second differentiation variable, bottom plots display disaggregated results. Error bars are 98\%-confidence bands over 100k paths.}\label{fig:gamma_results}
\end{figure}

\begin{figure}
	\centering
	\includegraphics[width=\figwidth]{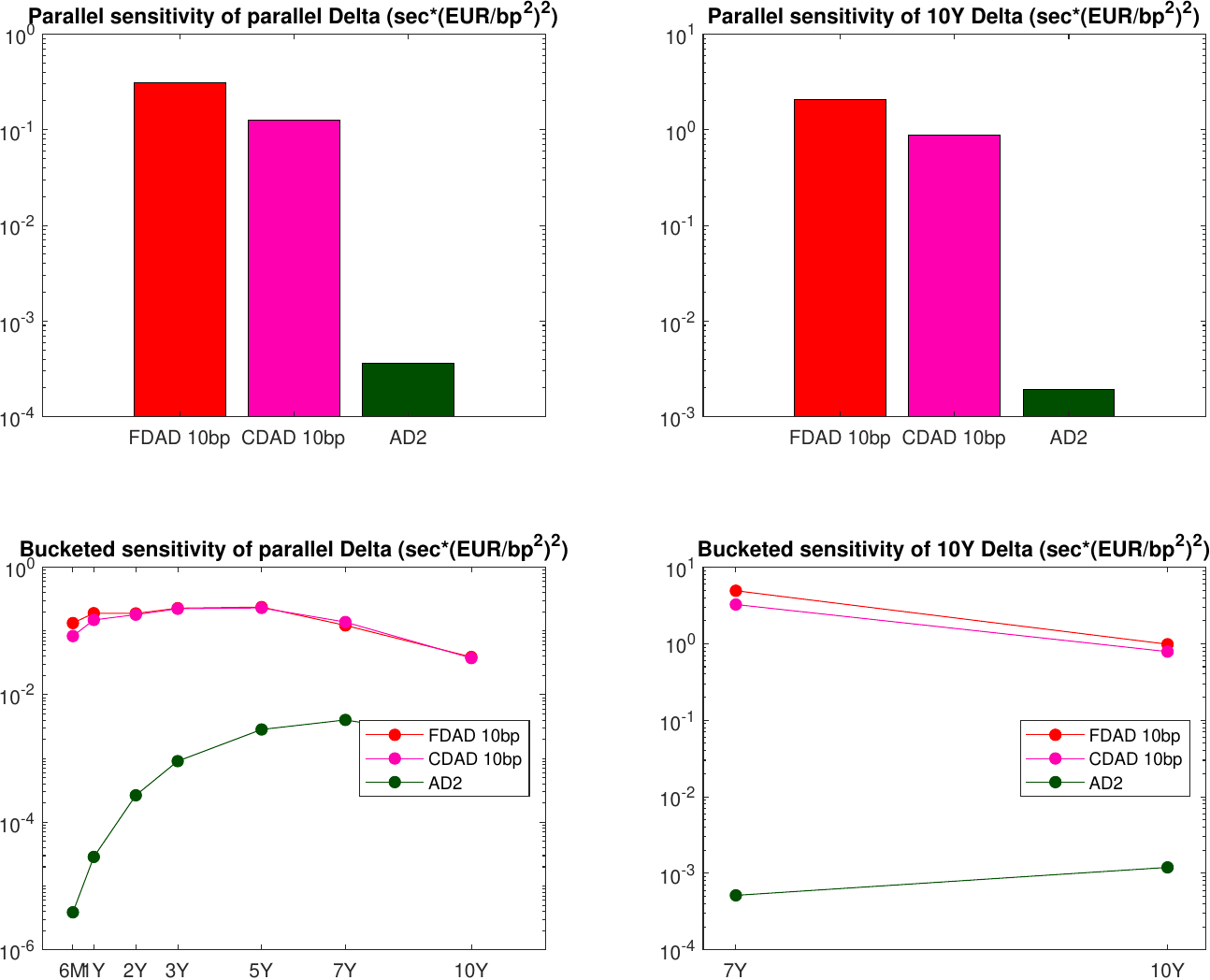}
	\caption{Credit Gamma efficiency, defined as the multiplication of the Monte Carlo variance of each sensitivity by the total computational time needed to compute the full credit-credit Hessian matrix. Graphs on the left refer to the aggregated derivatives along the first differentiation variable, while graphs on the right focus on the 10Y credit pillar. Top plots refer to sensitivities aggregated along the second differentiation variable, bottom plots to disaggregated results. The bottom-right plot includes only non-zero results. Note the log scale on the y axis.}\label{fig:gamma_efficiency}
\end{figure}

The same analysis was repeated for the second order derivatives where both differentiation variables are credit spreads. As expected, this is not the main convexity of CVA, with sensitivities of at most few Euros per basis point; as a consequence, only the very accurate AD2 is able to give statistically significant estimates with 100k Monte Carlo paths. This can be seen from the confidence bands of FD and CD in \cref{fig:gamma_results}, which approach or cross the horizontal axis in most cases. The bottom-right plot shows this phenomenon in the most extreme fashion: if one wants single pillar granularity on Credit Gamma, finite differences seem to produce pure noise, while AD2 is still remarkably accurate. Efficiency gains in \cref{fig:gamma_efficiency} reach three and more orders of magnitude at all levels of aggregation.

\section{Conclusion}\label{sec:conclusion}

The present paper shows that the main approaches to AAD for discontinuous payoffs can be adapted to the computation of CVA model Greeks up to the second order, and how to convert the results into sensitivities to market quotes.

In particular, a conditioning-based algorithm shows remarkable efficacy in two respects:
\begin{enumerate}
	\item It gives a very low cost-per-path estimator of all first and second order sensitivities where at least one differentiation variable explicitly affects only the credit model.
	\item Numerical experiments show that its results are also by far less noisy than more naive alternatives (still AAD-based).
\end{enumerate}	
The two above points combined imply order-of-magnitudes uncertainty-adjusted speed-ups, enough for almost real-time detailed first and second order CVA risk reports.

\section*{Acknowledgements}

Most of the first order results of \cref{sec:distributional,sec:conditional,sec:multiple}, while unpublished to date, were written some years ago under the supervision of Massimo Morini.
The whole paper has benefited from smart advice by Giorgio Facchinetti, and from useful user feedback by the traders Michele Trapletti, Marco Pinciroli and Francesco Mandelli.
The author would also like to thank the head of Interest Rates and Credit Models Nicola Moreni, and the head of Financial Engineering Andrea Bugin, for their support of this project.

\section*{Disclaimer}

The opinions expressed in this document are solely those of the author and do not represent in any way those of his employer.

\bibliography{biblio}

\begin{thebibliography}{}

\bibitem[Brigo et~al., 2013]{BrigoEtAl2013ctpyCollateralFunding}
Brigo, D., Morini, M., and Pallavicini, A. (2013).
\newblock {\em Counterparty Credit Risk, Collateral and Funding: With Pricing
  Cases for All Asset Classes}.
\newblock Wiley Finance.

\bibitem[Capriotti, 2015]{Capriotti2015secondOrder}
Capriotti, L. (2015).
\newblock Likelihood ratio method and algorithmic differentiation: Fast second
  order {G}reeks.
\newblock {\em Algorithmic Finance}, 4:81--87.

\bibitem[Capriotti et~al., 2011]{CapriottiEtAl2011realTimeCtpyCreditRiskInMC}
Capriotti, L., Lee, J., and Peacock, M. (2011).
\newblock Real time counterparty credit risk in {M}onte {C}arlo.
\newblock {\em Risk Magazine}, 24(86).

\bibitem[Chan and Joshi, 2015]{ChanJoshi2015optimalLimitMethods}
Chan, J.~H. and Joshi, M. (2015).
\newblock Optimal limit methods for computing sensitivities of discontinuous
  integrals including triggerable derivative securities.
\newblock {\em IEE Transactions}, 47:978--997.

\bibitem[Daluiso, 2016]{Daluiso2016modelToMarket}
Daluiso, R. (2016).
\newblock From model {G}reeks to market {G}reeks.
\newblock Available at SSRN: https://ssrn.com/abstract=2718194 or
  http://dx.doi.org/10.2139/ssrn.2718194.

\bibitem[Daluiso, 2020]{Daluiso2020secondOrder}
Daluiso, R. (2020).
\newblock Second-order {M}onte {C}arlo sensitivities in linear or constant
  time.
\newblock {\em Journal of Computational Finance}, 23(4):61--91.

\bibitem[Daluiso and Facchinetti, 2018]{DaluisoFacchinetti2018adDiscontinuous}
Daluiso, R. and Facchinetti, G. (2018).
\newblock Algorithmic differentiation for discontinuous payoffs.
\newblock {\em International Journal of Theoretical and Applied Finance}, 21.

\bibitem[Genz, 2004]{Genz2004numerical}
Genz, A. (2004).
\newblock Numerical computation of rectangular bivariate and trivariate normal
  and t probabilities.
\newblock {\em Statistics and Computing}, 14(3):251–--260.

\bibitem[Giles, 2007]{Giles2007mcSens}
Giles, M. (2007).
\newblock {M}onte {C}arlo evaluation of sensitivities in computational finance.
\newblock Technical Report NA07/12, Oxford University Computing Lab.

\bibitem[Glasserman, 2004]{Glasserman2004mcMethods}
Glasserman, P. (2004).
\newblock {\em {M}onte {C}arlo Methods in Financial Engineering}.
\newblock Springer, Berlin.

\bibitem[Griewank and Walther, 2008]{GriewankWalther2008evaluatingDerivatives}
Griewank, A. and Walther, A. (2008).
\newblock {\em Evaluating Derivatives: Principles and Techniques of Algorithmic
  Differentiation}.
\newblock Society for Industrial and Applied Mathematics, Philadelphia, second
  edition.

\bibitem[Henrard, 2011]{Henrard2011ADimplicitFunction}
Henrard, M. (2011).
\newblock Adjoint algorithmic differentiation: Calibration and implicit
  function theorem.
\newblock OpenGamma Quantitative Research n.1.

\bibitem[Henrard, 2013]{Henrard2013ADleastSquare}
Henrard, M. (2013).
\newblock Algorithmic differentiation in finance: Root finding and least square
  calibration.
\newblock OpenGamma Quantitative Research n.7.

\bibitem[Hull and White, 1990]{HullWhite1990pricingIRderivatives}
Hull, J. and White, A. (1990).
\newblock Pricing interest-rate derivative securities.
\newblock {\em The Review of Financial Studies}, 3(4):573--592.

\bibitem[Joshi and Zhu, 2016]{JoshiZhu2016optimalPartialProxyGammas}
Joshi, M. and Zhu, D. (2016).
\newblock Optimal partial proxy method for computing gammas of financial
  products with discontinuous and angular payoffs.
\newblock {\em Applied Mathematical Finance}, 23(1):22--56.

\bibitem[Morini, 2011]{Morini2011modelRisk}
Morini, M. (2011).
\newblock {\em Understanding and Managing Model Risk. A Practical Guide for
  Quants, Traders and Validators}.
\newblock Wiley Finance.

\bibitem[Pag\`{e}s et~al., 2016]{PagesEtAl2016adHigherOrder}
Pag\`{e}s, G., Pironneau, O., and Sall, G. (2016).
\newblock Vibrato and automatic differentiation for high order derivatives and
  sensitivities of financial options.
\newblock arXiv:1606.06143v1.

\end{thebibliography}

\end{document}

% The arXiv submission guidelines at https://trevorcampbell.me/html/arxiv.html and linked by arXiv itself suggest the following
% to trick arXiv into compiling the source 4 times, in order to ensure that autonum and cleveref references resolve properly:
\typeout{get arXiv to do 4 passes: Label(s) may have changed. Rerun}